\documentclass[prl,%showpacs
,twocolumn]{revtex4}
\usepackage{graphicx}
\newcommand{\be}{\begin{equation}} 
\newcommand{\ee}{\end{equation}} 
\newcommand{\bea}{\begin{eqnarray}} 
\newcommand{\eea}{\end{eqnarray}}

\usepackage{graphicx}
\usepackage{bm}

\begin{document}
\title{An Elementary Derivation of the Harmonic Oscillator Propagator}
\author{L. Moriconi}
\affiliation{Instituto de F\'\i sica, Universidade Federal do Rio de Janeiro \\C.P. 68528, Rio de Janeiro, RJ --- 21945-970, Brasil}
\begin{abstract}
The harmonic oscillator propagator is found straightforwardly from the free particle propagator, within the imaginary-time Feynman path integral formalism. The derivation presented here is extremely simple, requiring only elementary mathematical manipulations and no clever use of Hermite polynomials, annihilation \& creation operators, cumbersome determinant evaluations or any kind of involved algebra. 
\end{abstract}
%\pacs{}
\maketitle 
It is a current opinion that evaluations of the simple harmonic oscillator propagator are somewhat tricky. In fact, up to the author's knowledge, the usual arguments \cite{sakurai,holstein} confirm this belief to some extent. Our aim in this short note is to establish an alternative derivation of the euclidean harmonic oscillator propagator, which is technically very elementary. All one needs to know is the standard expression for the imaginary-time free particle propagator,
\bea
&&Z_{FP}=\int Dx \exp[- \int_0^T dt \frac{m}{2} \dot x^2]  \nonumber \\
&&= \sqrt{\frac{m}{2 \pi T}} \exp[- \frac{m}{2 T} (x_T-x_0)^2] \ , \
\eea
where, in concrete terms, the functional integration measure is defined as
\be
Dx \equiv \lim_{\epsilon_i \rightarrow 0} \sqrt{\frac{m}{2 \pi \epsilon_N}}\prod_{i=1}^{N-1} \sqrt{\frac{m}{2 \pi \epsilon_i}} dx_i 
\ . \
\ee
In (2), the time interval $T$ has been sliced into $N$ pieces, with sizes $\epsilon_i \equiv t_{i}-t_{i-1}$. We are interested to compute the imaginary-time harmonic oscillator propagator,
\be
Z_{HO}=\int Dx \exp[- \int_0^T dt (\frac{m}{2} \dot x^2+\frac{m \omega^2}{2} x^2)] \ . \
\ee
Observe that (3) can be rewritten as
\bea
&&Z_{HO}= \int Dx \exp \{- \int_0^T dt [ \frac{m}{2} (\dot x+\omega x)^2-\frac{m \omega}{2} \frac{d x^2}{dt}] \} \nonumber \\
&&= \exp[\frac{m \omega}{2} (x_T^2-x_0^2)] \nonumber \\
&& \times \int Dx \exp[- \int_0^T dt \frac{m}{2} (\dot x+\omega x)^2] \ . \
\eea
Taking $x(t) \equiv z(t) \exp(-\omega t)$ in (4), we get
\bea
&&Z_{HO}= \exp[\frac{m \omega}{2} (x_T^2-x_0^2)] [\prod_{i=1}^{N-1} \exp(-\omega t_i)] \nonumber \\
&& \times \int Dz \exp[- \int_0^T dt \frac{m}{2} \exp(-2 \omega t) \dot z^2] \ . \
\eea
The exponential factor $\exp(-2 \omega t)$ which appears above can be absorbed by time reparametrization. Let $t^\star \equiv \exp(2 \omega t) / 2 \omega + c$, where $c$ is an unimportant arbitrary constant. We will have, from (5),
\bea
Z_{HO}= \exp[\frac{m \omega}{2} (x_T^2-x_0^2)+ \frac{\omega T}{2}]  \nonumber \\
\times \int D z^\star \exp[- \int_{t_a}^{t_b} d t^\star \frac{m}{2} (\dot z^\star)^2] \ , \
\eea
where we have defined $z^\star ( t^\star) = z(t)$, $t_a=1/2 \omega +c$, and $t_b=\exp(2\omega T)/2 \omega +c$. To find (6), it is necessary to take into account the fact that $\epsilon_i^\star \equiv t_i^\star - t_{i-1}^\star = [\exp(2 \omega t_i)-\exp(2 \omega t_{i-1})]/2 \omega = 
\epsilon_i \exp(2 \omega \bar t_i)$, where $\bar t_i \equiv ( t_i + t_{i-1})/2$ \cite{comment}, so that
\bea
&&[\prod_{i=1}^{N-1} \exp(-\omega t_i)] Dz \equiv \sqrt{\frac{m}{2 \pi \epsilon_N}} \prod_{i=1}^{N-1} \exp(-\omega t_i) \sqrt{\frac{m}{2 \pi \epsilon_i}} dz_i \nonumber \\
&&=\exp(\frac{\omega T}{2})  \sqrt{\frac{m}{2 \pi \epsilon_N \exp(2 \omega \bar t_N)}}\prod_{i=1}^{N-1} 
\sqrt{\frac{m}{2 \pi \epsilon_i \exp(2 \omega \bar t_i)}} dz_i
\nonumber \\
&&=\exp(\frac{\omega T}{2})  \sqrt{\frac{m}{2 \pi \epsilon_N^\star}}\prod_{i=1}^{N-1} \sqrt{\frac{m}{2 \pi \epsilon_i^\star}} dz_i^\star
= \exp(\frac{\omega T}{2}) Dz^\star \ . \ \nonumber \\
\eea
Applying the free-particle expression (1) in (6), we obtain
\bea
&&Z_{HO}= \exp[\frac{m \omega}{2} (x_T^2-x_0^2)+ \frac{\omega T}{2}] \sqrt{\frac{m}{2 \pi (t_b-t_a)}} \nonumber \\
&& \times \exp[- \frac{m}{2 (t_b-t_a)} (z^\star_{t_b}-z^\star_{t_a})^2] = \sqrt{ \frac{m \omega}{2 \pi \sinh(\omega T)}} \nonumber \\
&&\times \exp \{- \frac{m \omega}{2 \sinh(\omega T)}[ (x_T^2+x_0^2) \cosh(\omega T) -2 x_T x_0] \}
\ , \ \nonumber \\
\eea
which is precisely the imaginary-time harmonic oscillator propagator.

I thank M. Moriconi for interesting discussions. This work has been partially supported by FAPERJ.


\begin{thebibliography}{21}
\bibitem[{\citenamefont{comment}()}]{comment} The choice 
$\bar t_i = (t_i+t_{i-1})/2$ is indeed the only prescription 
that leads to a time translational invariant propagator.
\end{thebibliography}
\end{document}